Spin Electronics

# Controllable reset behavior in domain wall-magnetic tunnel junction artificial neurons for task-adaptable computation

Samuel Liu[1][***], Christopher H. Bennett[2][**], Joseph S. Friedman[3][*], Matthew J. Marinella[2][*], David Paydarfar[4], and Jean Anne C. Incorvia[1][**]

[1] Department of Electrical and Computer Engineering, The University of Texas at Austin, Austin, TX 78712, USA
[2] Sandia National Laboratories, Albuquerque, NM 87123, USA
[3] Department of Electrical and Computer Engineering, The University of Texas at Dallas, Richardson, TX 75080, USA
[4] Department of Neurology, The University of Texas at Austin, Austin, TX 78712, USA

*Senior Member, IEEE*
*\*\* Member, IEEE*
*\*\*\* Student Member, IEEE*

**Abstract**—Neuromorphic computing with spintronic devices has been of interest due to the limitations of CMOS-driven von Neumann computing. Domain wall-magnetic tunnel junction (DW-MTJ) devices have been shown to be able to intrinsically capture biological neuron behavior. Edgy-relaxed behavior, where a frequently firing neuron experiences a lower action potential threshold, may provide additional artificial neuronal functionality when executing repeated tasks. In this study, we demonstrate that this behavior can be implemented in DW-MTJ artificial neurons via three alternative mechanisms: shape anisotropy, magnetic field, and current-driven soft reset. Using micromagnetics and analytical device modeling to classify the Optdigits handwritten digit dataset, we show that edgy-relaxed behavior improves both classification accuracy and classification rate for ordered datasets while sacrificing little to no accuracy for a randomized dataset. This work establishes methods by which artificial spintronic neurons can be flexibly adapted to datasets.

## I. INTRODUCTION

Within the past decade, rapid growth in data volume and complexity has accelerated the need for alternatives to CMOS-driven von Neumann architecture that has dominated modern computing. Contemporary data intensive tasks often expose the problem of a memory wall, where computation and memory are separate and processes are executed sequentially [Wulf 1995]. This leads to significant delay and energy consumption, particularly in prediction, interpolation, and extrapolation tasks that deal with large and complex sets of data. Due to these difficulties, research efforts have been directed toward eliminating these bottlenecks through neuromorphic computing, which draws inspiration from the parallel processing and efficiency of biological neural systems, e.g. the human brain. Hardware implementation of neuromorphic computing requires the development of artificial structures analogous to biological systems: neurons that are interconnected through synapses that represent the strength of connectivity. Artificial neural networks (ANNs) have the potential to overcome the speed and energy efficiency limitations faced by von Neumann computing [Furber 2016].

A fundamental building block of an ANN is the artificial integrate-and-fire (IF) neuron [Burkitt 2006]. A biological neuron receives impulses from other neurons through a synaptic network and builds up a membrane potential (integration). When this membrane potential reaches a threshold, the neuron generates an action potential, or a voltage spike, which it then propagates to other connected neurons in the network (firing). Further, leaky integrate-and-fire (LIF) neurons incorporate both the aforementioned property and the tendency of biological neurons to gradually lose membrane potential (leak) after a period without stimulation. There are numerous examples of this behavior implemented in artificial neuromorphic computing platforms [Nahmias 2013, Hassan 2018, Cui 2020].

An additional behavior of many biological neurons, for example mammalian pyramidal neurons in the cortex and hippocampus, is a transition between edgy and relaxed states. Consequently, the depolarizing shift causes the cell to be at a lower threshold for generating additional action potentials (edgy), while a neuron that does not fire often requires a higher threshold to generate action potential (relaxed). This phenomenon, called afterdepolarization, is an important mechanism underlying neural oscillations associated with information processing and rhythmic motor functions in the mammalian central nervous system [Llinas 1988, Bean 2007]. Similar to biological neurons, this edgy-relaxed behavior can be beneficial for artificial neurons that are used for datasets that have a degree of repetitive order.

A promising candidate in implementing artificial neurons is the domain wall-magnetic tunnel junction (DW-MTJ) device [Alamdar 2020, Currivan-Incorvia 2016, Currivan 2012], which consists of a perpendicularly magnetized ferromagnetic (FM) track containing a single DW and a sensing MTJ consisting of a fixed FM layer separated from the track by a thin insulating layer, depicted in Fig. 1. The DW propagates along $\hat{x}$ when a current is applied across the DW track through spin transfer torque (STT). If a heavy metal layer is introduced, the DW can also be propagated through spin orbit torque (SOT). The position of the DW represents the integration of the artificial neuron. The position of the sensing MTJ represents the firing threshold of the artificial neuron; as the DW passes underneath the sensing MTJ, a resistance change occurs, causing a spike in output current. The position of the DW is then reset in preparation for the next integration. The DW-MTJ device has been shown to implement IF behavior and exhibit energy efficiency both on the device and circuit levels [Sengupta 2016, Sharad 2012].

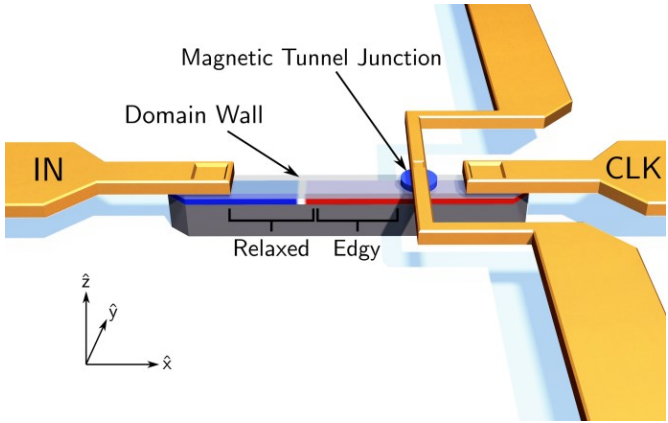

Fig. 1. Schematic of the DW-MTJ device. The red and blue in the CoFeB FM layer represent domains of antiparallel magnetization in $\hat{z}$. The white boundary represents the DW. The sensing MTJ is depicted as a blue disk aligned close to the right end of the track. The translucent layer represents the MgO insulating layer. The dark gray bottom layer depicts an optional heavy metal layer for SOT propagation. Generalized boundaries of edgy-relaxed states are also indicated. The convention for naming IN and CLK terminals is also shown.

Intrinsic magnetic properties such as shape anisotropy or a fixed external magnetic field have been shown to implement leaking behavior for DW-MTJs [Brigner 2019]; these attributes can also be exploited for edgy-relaxed behavior. This is accomplished by not resetting the neuron after firing, and instead allowing the DW to passively leak for a duration of time before integration begins again. If the time duration and the leaking velocity of the DW is tuned to partially reset the neuron, the result is that the recently fired neuron has a lower firing threshold than the other neurons in the layer. This soft reset mechanism has the added benefit of eliminating circuit overhead required for a current-driven hard reset signal.

In this work, we show using micromagnetic simulations and analytical modeling that by taking advantage of shape anisotropy and magnetic field to implement inherent edgy-relaxed behavior in DW-MTJ neurons, the delay and classification accuracy of an ANN is improved on data with repetitive inputs while sacrificing little to no accuracy for a completely randomized dataset. We also show that tuning the reset signal without device modifications to implement edgy-relaxed behavior can result in delay and classification accuracy improvements.

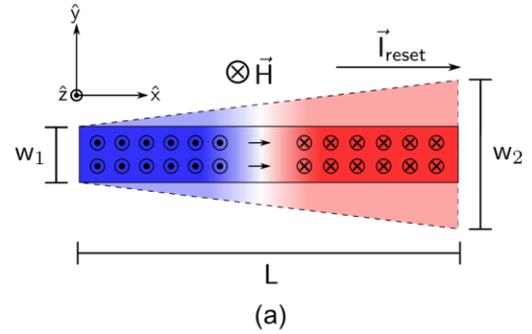

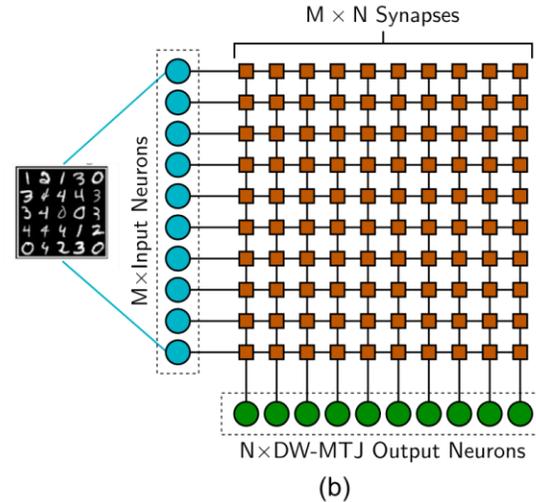

Fig. 2. (a) Top-down view of the dimensions of the FM layer of the DW-MTJ neuron. The DW is depicted in red. The soft reset methods using magnetic field, shape anisotropy, and reset currents are also shown. (b) Diagram of perceptron depicting input neurons and data, synapse network, and the simulated DW-MTJ output neurons. For Optdigits classification, there are 64 input neurons and 10 output neurons.

## II. METHODS

### A. Single Device Micromagnetics Model

To demonstrate the ability of a DW-MTJ neuron to implement edgy-relaxed behavior, a FM track containing a single DW is modeled using MuMax3 micromagnetics simulation software that solves the Landau-Lifshitz-Gilbert (LLG) equation [Vansteenkiste 2014]. We study the edgy-relaxed behavior for three possible leaking methods: magnetic field, shape anisotropy, and reset current, as defined in Fig. 2a. These reset options are compared against the baseline neuron that fully resets the DW to the left after firing.

The shape anisotropy-induced leaking is implemented on the model described in Brigner et al. [2019]. The simulated trapezoidal wire has dimensions 25 nm × 100 nm × 250 nm × 1.5 nm for $W_1 \times W_2 \times L \times$ thickness $t$, where $W_1$, $W_2$, and $L$ are defined in Fig. 2a. For all other reset methods, a rectangular wire of dimensions $W_1 = W_2 = 50$ nm, $L = 250$ nm, and t = 1.5 nm is used. The mesh size of the simulation was chosen to be 1 nm × 1 nm × 1.5 nm. The material parameters chosen are those of CoFeB: exchange stiffness $A_{ex} = 1.3 \times 10^{-11}$ J/m, saturation magnetization $M_{sat} = 0.8 \times 10^6$ A/m, uniaxial magneto-crystalline anisotropy in $\hat{z}$ $K_u = 5 \times 10^5$ J/m$^3$, Gilbert damping factor $\alpha = 0.05$, nonadiabicity factor $\xi = 0.05$, and spin polarization of $P = 0.7$. A temperature of 0K is assumed for the simulation.

### B. Neural Network and Classification Task

To quantify the utility of edgy-relaxed behavior in an ANN, a single layer neural network (perceptron) utilizing the edgy-relaxed DW-MTJs as output neurons is constructed. The task selected is classification of handwritten digits using the Optdigits dataset, a reduced resolution version of the MNIST handwritten digits dataset [Dua 2019]. To test the hypothesis that edgy-relaxed neurons excel in applications with datasets that have locality, a randomization scheme is devised. A test set of 100 images is sorted by digit into groups of 10. The completely sorted set is treated to have degree of randomness of 0. Randomness is introduced by randomly selecting $n$ images per digit group and swapping the images with a random counterpart drawn from the entire test set. The degree of randomness increases with $n$ until $n = 9$, where the dataset is fully shuffled.

The construction of the perceptron is tailored for classification of the Optdigits set. A set of 64 input neurons translate the pixel data into proportional voltage impulses. A network of 64 × 10 synapses is pre-trained using 900 images from the Optdigits set and treated as perfect weights. The 10 output neurons are simulated using an analytical model, where the index of the first neuron that fires is the classification of the image. A simplified diagram of the perceptron is shown in Fig. 2b. The baseline neuron classification accuracy of 87% and completion time of the test set of 1.14 µs were determined using 10 DW-MTJ output neurons. After classification of an image, the baseline neurons are hard reset using a 10 ns long current pulse. The reset circuitry is assumed to only be able to send minimum reset pulses of 10 ns duration for feasible implementation. The classification accuracy achieved is the theoretical accuracy possible for the pre-trained synapse array on the input data, independent of the magnetic devices.

### C. Analytical DW-MTJ Neuron Model

In order to evaluate the performance of edgy-relaxed DW-MTJ neurons, an analytical model is used to simulate the output neurons of the perceptron. A one-dimensional solution of DW motion is used as described by Beach et al. [2008]. Due to the observation that the DW in the shape anisotropy-driven soft reset case is in precession, the velocity of the DW in Walker breakdown regime is the following:

$$\bar{v} = \alpha\gamma\Delta H_{eff} + \frac{g\mu_B P}{2eM_{sat}}j \quad (1)$$

where γ is the gyromagnetic ratio, Δ is the width of the DW, $H_{eff}$ is the effective external magnetic field, $g$ is the Landé factor, $\mu_B$ is the Bohr magneton, $e$ is electron charge, and $j$ is current density applied in the $-\hat{x}$. The effect of the sloped shape of the DW is approximated to have the effect of a constant external field. This assumption is accurate when the DW-MTJ neuron has a slight exponential curve to the DW track instead of a straight slope [Brigner 2020]. The physical dimensions of the simulated analytical synapses are approximated to be the same as that of the device simulated in Section IIA.

For the cases of external field-driven and current-driven soft reset cases, the DW remains in the Néel configuration due to the straight track. As a result, the expression for DW velocity is for a DW in the non-Walker breakdown regime:

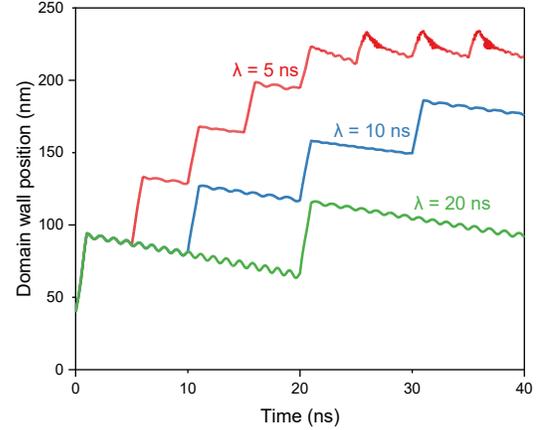

Fig. 3. DW position as a function of time as a result of repeated impulses for varying periods of time for the shape anisotropy neuron with soft reset.

$$\bar{v} = \frac{\gamma\Delta H_{eff}}{\alpha} + \frac{g\mu_B P}{2eM_{sat}}j \quad (2)$$

The velocity of the DW is calculated at each timestep based on input current density and effective field. The new position is then calculated relative to the DW position at the previous timestep. A fixed timestep of 0.2 ps was used for the simulation. The fully relaxed position of the DW is set to be $x = 25$ nm due to the inclusion of 10 nm fixed magnetization regions at the ends of the device and the approximate width of the DW.

## III. RESULTS

### A. Single Device Edgy-Relaxed Behavior

Figure 3 shows the inherent edgy-relaxed behavior of a single shape anisotropy-driven DW-MTJ neuron. Here, using the micromagnetic model, a 80 µA pulse with 1 ns duration is applied for electron flow from IN to CLK with periods (λ) of 5 ns (red), 10 ns (blue), and 20 ns (green). The DW position vs. time plot shows the periods of integration during the pulse duration, when the DW position increases as it moves right towards the output MTJ, as well as periods of leaking after the current pulse is removed and the DW moves back towards the left of the FM track. After e.g., 40 ns, the DW stimulated every 5 ns is farthest along the track, followed by 10 ns period and 20 ns period, respectively. Due to this, the neuron stimulated every 5 ns has the lowest firing threshold since it remains closer to the output MTJ, while the neuron stimulated every 20 ns has the highest firing threshold. Similar behavior is observed for the field-driven and current-driven reset methods (not shown). This demonstrates that with a passive leaking mechanism instead of a hard reset after firing, a DW-MTJ neuron can produce inherent behavior analogous to biological edgy-relaxed neurons.

### B. Shape Anisotropy-Driven Soft Reset

The utility of the edgy-relaxed behavior in DW-MTJ neurons in ANNs is addressed in this and subsequent sections for the different possible leaking methods. For shape anisotropy-driven leaking DW-

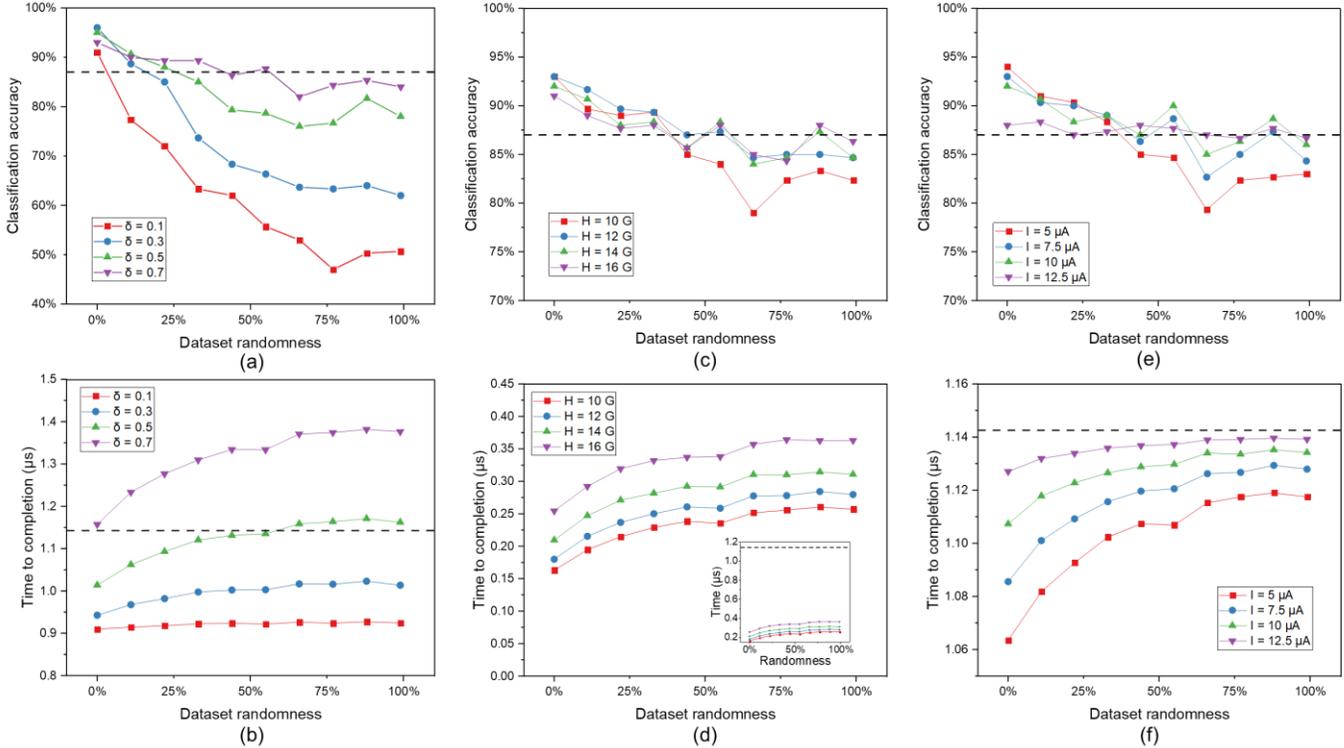

Fig. 4. Classification accuracy and completion time as a function of dataset randomness for varying (a) (b) device slopes $\delta$, (c) (d) external magnetic fields $H$, (e) (f) and reset current amplitudes $I$. The baseline network classification accuracy of 87% and completion time of 1.14 μs are shown as black dashed lines (inset for (d)).

MTJs, output neurons with varying slopes of $\delta = \frac{W_2 - W_1}{L} = 0.1, 0.3, 0.5,$ and $0.7$ are used to classify datasets of variable randomization using the analytical model of (1) and the crossbar array described in Section IIC. Because a racetrack with a larger width carries a current density penalty, we set $W_1 = 0$ nm to obtain competitive results; the geometry is that of a triangular track. Because the leaking effect of shape anisotropy is relatively weak for tracks with physically reasonable slopes, a rest with a period of 9 ns is implemented after each classification to allow soft reset of the DW to reduce classification errors otherwise seen in higher randomness datasets. From the result in Fig. 4a, for all geometries, there is enhanced classification accuracy for datasets with a relatively low degree of randomness. This improvement over the baseline is due to the repetition of presented images in each subgroup, a short-term learned repetition due to the edgy-relaxed neuron. The classification accuracy of a completely random dataset is also significantly increased with greater leak strength, at 50.67% for $\delta = 0.1$ versus 84.00% for $\delta = 0.7$, comparable to the performance of the baseline network. However, this geometry results in a significant time penalty.

Figure 4b depicts the total classification time of the test set. In general, for all geometries, the greatest reduction in completion time results when the edgy-relaxed neuron is applied to a non-random dataset and the completion time generally increases with increasing randomness. Additionally, the completion time increases with device slope due to the increased leaking effect along with the greater penalty to current density as the DW approaches the wider end of the device.

The advantage of this soft reset method is that this can be implemented by modifying the shape of the device and eliminates the need for reset circuitry. However, the leaking effect is weaker and results in penalties to classification accuracy and completion time.

### C. Magnetic Field-Driven Soft Reset

External magnetic fields of $H$ = 10 G, 12 G, 14 G, and 16 G are applied in $-\hat{z}$ on output neurons with rectangular dimensions 50 nm wide × 250 nm long × 1.5 nm thick and implemented in the ANN from Section IIC using (2). The range of magnetic field strengths were chosen to be achievable values for a permanent magnet in proximity to the DW-MTJ neuron that could be engineered near the device or into the thin film stack. Due to the much stronger leaking effect, a rest with a period of 2 ns is implemented after each classification to allow a soft reset of the DW.

In Fig. 4c, similar to the shape-based leaking, there is enhancement in classification accuracy for non-random datasets and there is less accuracy penalty for highly randomized datasets than for the shape anisotropy-driven case. It is also observed that as field strength increases, the more classification accuracy converges to the performance of the baseline neuron.

Due to field-induced DW leak strength, there is a significant reduction in completion time, shown in Fig. 4d. All field strengths resulted in significantly reduced completion time compared to the baseline neuron. The completion time in general increases with dataset randomness as well as field strength, though all results are well below the completion time of the baseline neuron.

The magnetic field implementation of edgy-relaxed behavior results in significant performance improvements. In the case of classification time, a strong enough magnetic field can provide a 3-6x speedup depending on dataset order. In addition, classification accuracy can be improved by around 5% for ordered datasets while there is no significant loss in accuracy for random datasets. In addition, this can be implemented into an intrinsic device property through the placement of a permanent magnet near the device, which may reduce the overhead that results from reset circuitry. However, due to the field-induced DW leaking strength, there is a risk of higher energy dissipation needed to propagate the DW in the integration direction.

### D. Reset Signal Amplitude-Driven Soft Reset

The effect of a current-delivered soft reset without a leaking mechanism is studied by applying reset currents $I_{reset}$ = 5 µA, 7.5 µA, 10 µA, and 12.5 µA to DW-MTJ neurons with rectangular dimensions. The self-imposed limitation of a minimum 10 ns duration reset pulse is also applied in this case.

Figure 4e shows that like the other methods discussed, a current-driven soft reset also enhances classification accuracy for non-random datasets over that of the baseline neuron. This holds until around 30% randomness. In general, the greater the magnitude of the reset current pulse, the more the classification accuracy converges to the result of the baseline neuron. For the $I_{reset}$ = 12.5 µA case, the reset current pulse is close enough to a hard reset that the resulting trend is converging with the baseline neuron result and the enhancement to classification accuracy for the datasets with locality is largely lost.

From Fig. 4f, there is an improvement in completion time for all reset current amplitudes, though there is no result that has a greater than 10% reduction. This is likely due to the limitation that the reset signal must have a minimum of 10 ns pulse duration. Though variation is not very large, the completion time generally increases with dataset randomness and reset current amplitude.

While this method comes with reset circuitry overhead, it does not require added device complexity. Additionally, this implementation of edgy-relaxed behavior can also be applied to other devices, allowing for performance improvements while not significantly changing the overall design of the device. Another advantage of current-driven soft reset is reconfigurability; it can be tuned dynamically using software. For example, an algorithm can be used to determine prevalence of repeated data and adjust the strength of edgy-relaxed behavior accordingly.

## IV. CONCLUSIONS

Biological neurons have an adaptive edgy-relaxed transition behavior when repeatedly stimulated, leading to a reduced threshold for action potential. This work demonstrates ways this behavior can be implemented intrinsically in a DW-MTJ LIF neuron by exploiting magnetic properties of shape anisotropy and external magnetic field to manipulate DW movement. An additional method of modulating the reset pulse is also proposed and can be widely applied to other types of analog neuron devices. By modulating the strength of the leak and degree of soft reset, the classification accuracy on a dataset with repeated data can be improved significantly due to the intrinsic short-term memory of edgy-relaxed behavior. Additionally, there is a significant improvement in classification time for external field implementation and a slight improvement for reset pulse modulation. This work establishes ways that unique magnetic properties can be utilized to implement edgy-relaxed behavior in a DW-MTJ neuron by incorporating these effects into device design, which has the potential to reduce the overhead required by reset circuitry. These results can lead to ANNs that can be adapted to different types of expected datasets through modification of neuron device behavior.

## ACKNOWLEDGMENT

The authors acknowledge funding from the National Science Foundation CAREER under award number 1940788, funding from the Sandia National Laboratories Laboratory Directed Research and Development (LDRD) Program, and computing resources from the Texas Advanced Computing Center (TACC) at the University of Texas at Austin (http://www.tacc.utexas.edu). This paper describes objective technical results and analysis. Any subjective views or opinions that might be expressed in the paper do not necessarily represent the views of the U.S. Department of Energy or the United States Government. Sandia National Laboratories is a multimission laboratory managed and operated by NTESS, LLC, a wholly owned subsidiary of Honeywell International Inc., for the U.S. Department of Energy's National Nuclear Security Administration under contract DE-NA0003525.